\documentclass[fleqn,usenatbib]{mnras}

\usepackage{newtxtext,newtxmath}

\usepackage[T1]{fontenc}
\usepackage{ae,aecompl}

\usepackage{graphicx}	
\usepackage{dblfloatfix}
\usepackage{amsmath}	
\usepackage{threeparttable}
\usepackage{subcaption}
\usepackage{caption}
\usepackage{color}
\usepackage[dvipsnames]{xcolor}
\usepackage[normalem]{ulem}
\usepackage{booktabs}

\newcommand{\fermi}{\textit{Fermi}-{\rm LAT}}
\newcommand{\planck}{\textit{Planck}}

\newcommand{\gray}{$\gamma$-ray}
\newcommand{\grays}{$\gamma$-rays}
\newcommand{\xray}{$\rm X$-ray}

\newcommand{\msun}{\mbox{$M_\odot$}}

\def\deg{\hbox{$^\circ$}}

\title[\fermi\ detection around RCW38]{GeV \gray\ emission in the field of young massive star cluster RCW 38}
\author[Ge et. al]{Ting-Ting Ge$^{1}$, 
Xiao-Na Sun$^{1}$\thanks{E-mail:xiaonasun@gxu.edu.cn}, 
Rui-Zhi Yang$^{2}$$^{3}$$^{4}$,
Pak-Hin Thomas Tam$^{5}$$^{6}$,
Ming-Xuan Lu$^{1}$
\newauthor{En-Wei Liang$^{1}$}
\\
$^{1}$Guangxi Key Laboratory for Relativistic Astrophysics, School of Physical Science and Technology, Guangxi University, Nanning 530004, China\\
$^{2}$Department of Astronomy, School of Physical Sciences, University of Science and Technology of China, Hefei, Anhui 230026, China\\
$^{3}$CAS Key Labrotory for Research in Galaxies and Cosmology, University of Science and Technology of China, Hefei, Anhui 230026, China\\
$^{4}$School of Astronomy and Space Science, University of Science and Technology of China, Hefei, Anhui 230026, China\\
$^{5}$School of Physics and Astronomy, Sun Yat-sen University, Zhuhai 519082, China\\
$^{6}$CSST Science Center for the Guangdong-Hong Kong-Macau Greater Bay Area, Sun Yat-Sen University, Zhuhai 519082, China\\}

\pubyear{2023}

\begin{document}
\label{firstpage}
\pagerange{\pageref{firstpage}--\pageref{lastpage}}
\maketitle

\begin{abstract}
We report the detection of \gray\ emission by the Fermi Large Area Telescope (\fermi) towards the young massive star cluster RCW 38 in the 1-500 GeV photon energy range.
We found spatially extended GeV emission towards the direction of RCW 38, which is best modelled by a Gaussian disc of 0.23$\deg$ radius with a significance of the extension is $\sim 11.4 \sigma$. 
Furthermore, the spatial correlation with the ionized and molecular gas content favors the hadronic origin of the \gray\ emission. The \gray\ spectrum of RCW 38 has a relatively hard photon index of $2.44 \pm 0.03$, which is similar to other young massive star clusters. We argue that the diffuse GeV \gray\ emission in this region likely originates from the interaction of accelerated protons in the stellar cluster with the ambient gas.

\end{abstract}




\section{Introduction}
Young massive star clusters (YMCs), which can drive high-speed stellar winds to speeds up to thousands of kilometres per second by radiation pressure of the OB stars, are believed to be 
potential accelerators of Galactic cosmic rays (CRs) \citep{1982Abbott, 1983Cesarsky} . 
Winds from star clusters provide a suitable environment for particle acceleration. Since the high-speed winds exist for $\sim 10^{6}$ years, which is much longer than the duration of the supernova remnant (SNR) shock, it is possible that the massive stars in some clusters are more efficient accelerators than SNRs \citep{2020Bykov}. Observationally, several YMCs have been detected in the GeV or TeV \gray\ band, most of which have significant spatial extensions up to more than 50 pc and a hard \gray\ spectrum that can be described by a power-law function with an index of 2.2--2.4. More interestingly, a derived $1/r$ profile of the CR spatial distribution would indicate a continuous CR injection process of a steady central source in some cases \citep{2019Aharonian}.
\fermi\ has detected diffuse GeV \gray\ emission around some massive star clusters, such as Cygnus Cocoon \citep{2011Ackermann}, NGC 3603 \citep{2017Yang}, Wersterlund 1 \citep{2012Abramowski}, Westerlund 2 \citep{2018Yang}, RSGC 1 \citep{RSGC1Sun}, W40 \citep{W40Sun}, NGC 6618 \citep{2022Liu}, and Carina \citep{2022Ge}.

%
RCW 38 is one of the few young massive star clusters in our Galaxy, with an age of less than 1 Myr \citep{2006Wolk,2011Winston}. Its young age makes it particularly relevant for star formation studies \citep{2022Ascenso}. RCW 38 has large total mass, similar to that of the Orion Nebula Cluster, and is an active high-mass star-forming region, the second nearest to the Sun, at a distance of 1.7 kpc \citep{2003Lada, 2006Wolk, 2008Wolk}.
In the central part of RCW 38, two remarkable infrared peaks have been identified at 2 $\mu$m  and 10 $\mu$m, respectively \citep{1974Frogel}. It contains $\sim$ 10,000 stars \citep{2015Kuhn}, and 30 OB star candidates have been identified in the RCW 38 cluster \citep{2006Wolk, 2011Winston}, of which 20 OB stars are within $\sim$ 0.5 pc of the cluster centre \citep{2006Wolk}. At the centre of the RCW 38 cluster, a massive binary of spectral type $\sim$ O5 \citep{2009DeRose} ionized the region \citep{2008Wolk}, a dense molecular cloud envelopes the bright \ion{H}{II} region \citep{2023Fukushima}. Massive stars born in such a region would have a strong effect on their surroundings through stellar winds and supernova explosions. 
For RCW 38, diffuse \xray\ emission from this region has been observed by $Chandra$ and Suzaku \citep{2002Wolk,2023Fukushima}. \cite{2002Wolk} find the non-thermal \xray\ emission from the embedded massive star-forming region RCW 38. 

Molecular line observations towards RCW 38 are very limited. Two molecular clouds associated with the RCW 38 cluster were first detected in CO at a velocity of 1 $\rm km\ s^{-1}$ \citep{1979Gillespie}. \cite{1999Yamaguchi} observed a large region of RCW 38 in the CO emission with NANTEN and estimated its mass to be $1.5 \times 10^{4}\ \msun$. \cite{2016Fukui} observed a new molecular line towards the RCW 38 cluster, obtained with the NANTEN2, Mopra and ASTE telescopes, showing two molecular clouds with velocities of 2 and 14 $\rm km\ s^{-1}$, and the total mass of the clouds and the cluster is less than $10^{5} \msun$. The 2 $\rm km\ s^{-1}$ cloud has a ring-like shape with a cavity ionized by this cluster and has a high molecular column density of $\sim 10^{23}\ \rm cm^{-2}$ \citep{1995Zinchenko, 1999Yamaguchi, 2016Fukui}. The other one is the finger cloud (the 14 $\rm km\ s^{-1}$ cloud), which has a tip towards the cluster \citep{2008Gyulbudaghian,2016Fukui}.

%
Another complicating factor in studying RCW 38 is that it lies near the Vela Molecular Cloud Ridge (VMR). VMR identified by \cite{1991Murphy} is a giant molecular cloud complex located in Vela and Puppis. Potential particle acceleration sites in the VMR include star-forming region, the expansive \ion{H}{II} Gum Nebula, the Vela SNR, and the shell-type SNR RX J0852.0-466 \citep{2018Prisinzano, 2018HESS, 2019Massi}. Recently, \cite{Peron:2023nu} analyzed \fermi\ data in the VMR region and found significant \gray\ emission extending to a few tens of GeV, in correspondence of several \ion{H}{II} regions. They concluded that the \gray\ emission could be connected to \ion{H}{II} regions, which are associated with stars including the two most bright \gray\ sources of their sample, RCW38 and RCW36.

Powerful shocks of the massive stars in YMC RCW 38 produced by the interaction of their stellar winds interacting with the interstellar medium, are likely to accelerate particles to very high energy \citep{2012Del, 2021Morlino}. For the released \fermi\ 12-yr Source Catalogue (4FGL-DR3) \citep{2020Abdollahi, 2022Abdollahi}, a \gray\ point source 4FGL J10859.2-4729 has been detected in the direction of the RCW 38 region. For such a complex region described above, the origin of these \grays\ remains a mystery that requires careful and comprehensive investigation. In this paper, we have analyzed the \gray\ emission towards RCW 38, taking advantage of more than 14 years of \fermi\ data, and tried to investigate the possible origin of the RCW 38 \gray\ emission. This paper is organized as follows. In Sect.\ref{sec:data}, we present the data set and the results of the data analysis. In Sect.\ref{sec:Gas}, we study the gas distributions in this region. In Sect.\ref{sec:origin}, we investigate the possible origin of the GeV \gray\ emissions. Finally, we discuss the implications of our study in Sect.\ref{sec:Conc} and summarize the main conclusions in Sect.\ref{sec:Disc}.

\section{\fermi\ data analysis}
\label{sec:data}
We selected the latest \fermi\ Pass 8 data around the RCW 38 region from August 4, 2008 (MET 239557417) until October 26, 2022 (MET 688445051) and used the standard LAT analysis software package $\it{v11r5p3}$\footnote{\url{https://fermi.gsfc.nasa.gov/ssc/data/analysis/software/}}. 
We chose a radius of 10$\deg$ centred at the position of RCW 38 (R.A. = 134.77$\deg$, Dec. = - 47.51$\deg$) as the region of interest (ROI).
The instrument response functions (IRFs) {\it P8R3\_SOURCE\_V3} was selected to analyze the events in the ROI of evtype = 3 and evclass = 128. 
We also applied the recommended expression $\rm (DATA\_QUAL > 0) \&\& (LAT\_CONFIG == 1)$ to select the Good Time Intervals (GTIs) based on the information provided in the spacecraft file.
In order to reduce the \gray\ contamination from the Earth's albedo, only the events with zenith angles less than 100$\deg$ are included in the analysis. We used the Python module \footnote{\url{https://fermi.gsfc.nasa.gov/ssc/data/analysis/scitools/python_tutorial.html}} which implements a maximum likelihood optimization technique for a standard binned analysis.

In the background model, we included the recently released \fermi\ 12-year Source Catalog of point-like and extended sources (4FGL-DR3, \cite{2020Abdollahi,2022Abdollahi}) within a region of a radius 15$\deg$ around RCW 38. 
The source model file was generated using the script make4FGLxml.py\footnote{\url{https://fermi.gsfc.nasa.gov/ssc/data/analysis/user/}}, and parameters of all sources within 5$\deg$ of center were allowed to vary.
For the diffuse background components, we use the latest Galactic diffuse emission model {\it gll\_iem\_v07.fits} and the isotropic extragalactic emission model {\it iso\_P8R3\_SOURCE\_V3\_v1.txt}\footnote{\url{https://fermi.gsfc.nasa.gov/ssc/data/access/lat/BackgroundModels.html}}, allowing their normalization parameters to vary.

\begin{figure}
\includegraphics[scale=0.35]{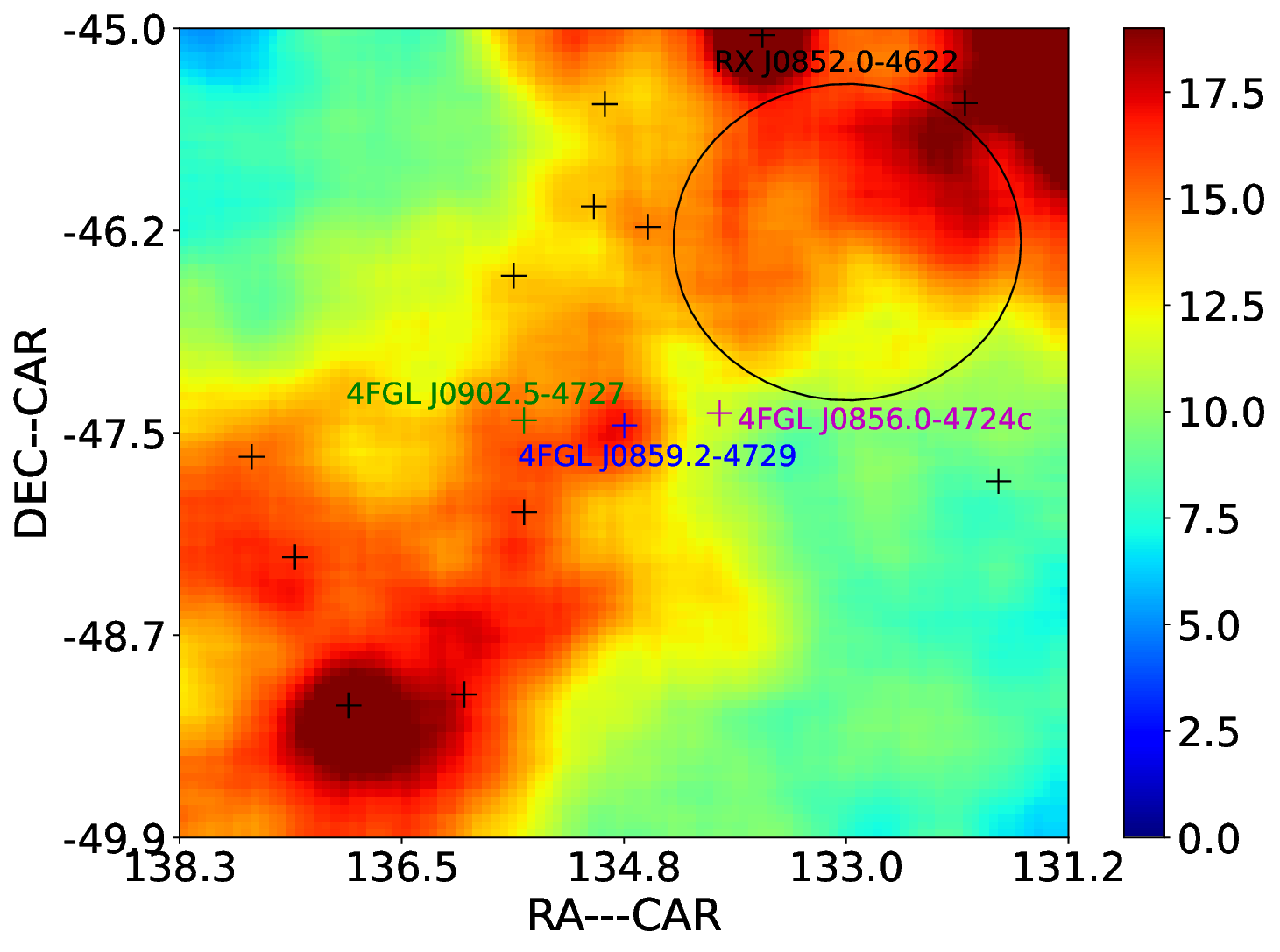}
\caption {\fermi\ counts map above 1 GeV in the $5^{\deg} \times 5^{\deg}$ region around RCW 38, with a pixel size of $0.05^{\deg} \times 0.05^{\deg}$. The black circle shows the extended emission related to RX J0852.0-4622 \citep{2011Tanaka}. The coloured crosses indicate the three point sources 4FGL J0859.2-4729 , 4FGL J0902.5-4727, 4FGL J0856.0-4724c around RCW 38. Black crosses represent other 4FGL-DR3 sources within the region.}
\label{fig:cmap}
\end{figure}

\subsection{Spatial analysis}
First, we used the events above 1 GeV to study the spatial distribution of the \gray\ emission near RCW 38. The \gray\ counts map in the $5\deg \times 5\deg$ region around RCW 38 is shown in Fig.\ref{fig:cmap}. 
We note that there are three 4FGL-DR3 catalog point sources 4FGL J0859.2-4729 , 4FGL J0902.5-4727, 4FGL J0856.0-4724c around RCW 38.  

\subsubsection{Single point source model}
\label{sec:model1}
To study the excess GeV \gray\ emission around RCW 38, we test several different models. The models tested are summarized in Table \ref{table:1}. 
We excluded the unidentified sources 4FGL J0859.2-4729, 4FGL J0902.5-4727, and 4FGL J0856.0-4724c from the background model. We performed a background-only fitting and generated the residual test statistic (TS) map shown in Fig.\ref{fig:residual}. We assumed that the \gray\ excess within the RCW 38 cluster arises from a single component. We added a point-like source at the cluster position into our background model and optimized the localization using the {\it gtfindsrc} tool. 
The best-fit position of this point source above 1 GeV is [RA = $134.795\deg$ , Dec = $-47.514\deg$] with $2\sigma$ error radius of $0.021\deg$, marked with white crosses p1 in Fig.\ref{fig:residual}. RCW 38 is $0.015\deg$ away from the best-fit position, within the $2\sigma$ error circle (consistent with the containment angle) \citep{2011Winston}. 
Our first model (model 1) uses a single point source (p1) with a LogParabola spectral shape to model the excess \gray\ emission around RCW 38 region. 
We performed a binned likelihood analysis to derive the likelihood value ($-\log({\cal L})$) and the Akaike Information Criterion (AIC, \cite{Akaike1974}) value. The AIC is defined as AIC = $-2\log({\cal L}) + 2k$, where $k$ is the number of free parameters in the model. 
The derived $-\log({\cal L})$ and AIC for the single source model are -3498059 and -6995952, respectively.

\subsubsection{Spatial template for Gaussian disc}
\label{sec:model2}
As shown in Fig.\ref{fig:residual}, the residual \gray\ emission in this region is diffuse. Therefore, we considered the Gaussian discs or uniform discs templates to test whether the residual emission is extended. The centre of the discs are set at the best-fit position with various radii from $0.1\deg$ to $0.6\deg$ in steps of $0.05\deg$.
We used these Gaussian discs or uniform discs to replace the spatial components of the single point source in the model 1. Then, we fitted the observations using the above modified models and calculated the corresponding significance of the source extension $\rm TS_{ext}$. $\rm TS_{ext}$ is quantified by $\rm TS_{ext}=2\log({\cal L}_{ext} /{\cal L}_{ps})$, where $\rm {\cal L}_{ext}$ is the maximum likelihood for the extended source model, and $\rm {\cal L}_{ps}$ for the single point source model \citep{2012Lande}. 
In the above test, we find that the Gaussian disc with a radius of $0.35\deg \pm 0.05$ (model 2) is preferred than the other models and the derived $\rm TS_{ext}$ is 82, meaning the significance of the extension is about $9.1\sigma$. The derived $-\log({\cal L})$ and AIC for model 2 are -3498100 and -6996034 after performing the binned likelihood analysis. 

\subsubsection{Spatial template for molecular and ionized hydrogen}
\label{sec:model3}
To determine whether the extended GeV \gray\ emission is correlated with the gas distribution, we considered a spatial template of molecular hydrogen (H$_{2}$) and ionized hydrogen (\ion{H}{II}). Because the \gray\ emission has a good spatial correlation with both the H$_{2}$ and \ion{H}{ii} gases, especially the \ion{H}{ii} gas (see Sect.\ref{sec:Gas} and Fig.\ref{fig:Gas}). Thus, we summed the column densities of the H$_{2}$ and \ion{H}{ii} gases to generate the H$_{2}$ and \ion{H}{ii} template. We assumed that the gas template has a LogParabola spectral shape. Then, we added the H$_{2}$ + \ion{H}{ii} template (model 3) to replace the spatial components of the single point source in the model 1. After performing the binned likelihood analysis, the derived $-\log({\cal L})$ and AIC for model 3 are -3498091 and -6996016, respectively. In addition, we also adopted a spatial template considering \ion{H}{II} gas and added into background model as our model 4. The derived $-\log({\cal L})$ and AIC for model 4 are -3498113 and -6996060 after performing the binned likelihood analysis.

\subsubsection{Three point sources model}
\label{sec:model4}
\begin{figure}
\includegraphics[scale=0.35]{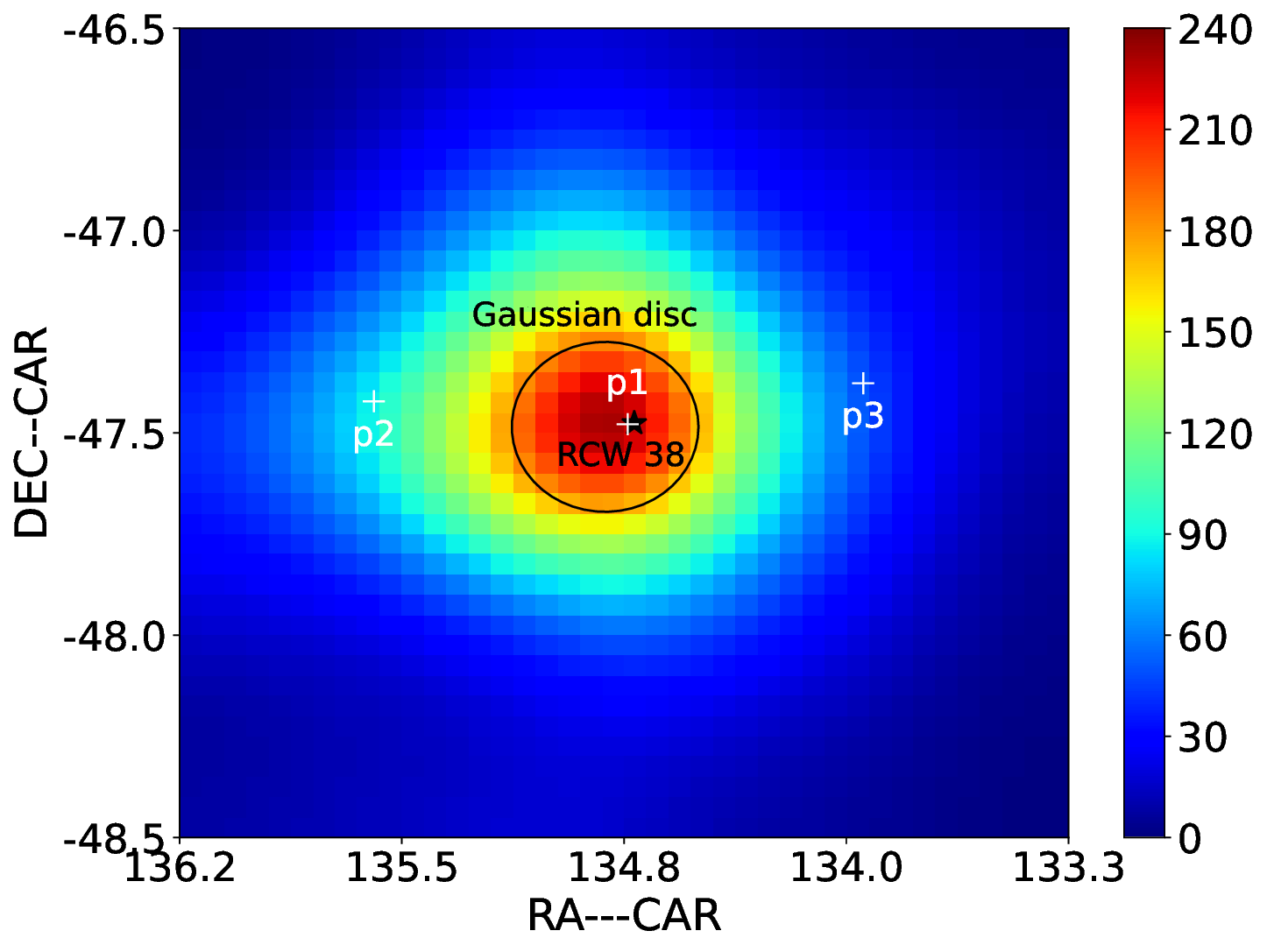}
\caption {\fermi\ TS residual map above 1 GeV near the RCW 38 after excluding the sources 4FGL J0859.2-4729, 4FGL J0902.5-4727, and 4FGL J0856.0-4724c from the background model. Details are given in Sect. \ref{sec:model1}. The map has a size of $2\deg \times\ 2\deg$ ($0.05\deg \times\ 0.05\deg$ pixel size) and has been smoothed with a Gaussian kernel of 0.45$\deg$. The black asterisk indicates the infrared position of RCW 38 \citep{2011Winston}, and the black circle marks the Gaussian disc with a radius of $0.23\deg$. The "p1" symbol shows the best-fit position of the assumed single point source used for the spatial analysis. The "p2" and "p3" indicate the point sources 4FGL J0902.5-4727 and 4FGL J0856.0-4724c, respectively, listed in the 4FG-DR3 catalog.}
\label{fig:residual}
\end{figure}

In model 1, we removed the two point sources 4FGL J0902.5-4727 and 4FGL J0856.0-4724c close to the RCW 38 cluster from the background model.
4FGL J0902.5-4727 and 4FGL J0856.0-4724c are marked with white crosses p2 and p3 in Fig.\ref{fig:residual}. Note that p2 and p3 are approximately 0.54$\deg$ and 0.57$\deg$ from the centre of RCW 38, respectively. Based on model 1, we added the two 4FGL-DR3 point sources back into the model file (model 4). Each component has a power-law spectral shape. Through the performing of the binned likelihood analysis, the derived $-\log({\cal L})$ and AIC for the three point source model are -3498105 and -6996036, respectively.

\begin{table*}
        \caption{Spatial analysis (> 1 GeV) results for the different models.}
\begin{tabular}{lcccc}
\hline
\hline
        model & -$\log({\cal L})$ & $\rm TS_{ext}$ & d.o.f. & $\rm \Delta AIC$ \\

\hline
        model 1 (single point source) & -3498059 & - & 83 & - \\ 
        model 2 ($0.35\deg$ Gaussian disc) & -3498100 & 82 & 83 & -82 \\ 
        model 3 ($\rm H_2$ + \ion{H}{ii} template) & -3498091 & 64 & 83 & -64 \\ 
        model 4 (\ion{H}{ii} template)     &  -3498113   & 108 & 83 & -108   \\ 
        model 5 (p1 + p2 + p3) & -3498105 & - & 87 & -84 \\ 
        model 6 ($0.23\deg$ Gaussian disc + p2 + p3) & -3498124 & 130 & 87 & -122 \\ 
\hline
\hline
\end{tabular}
\label{table:1}
\end{table*}

\subsubsection{Spatial template for two point sources and Gaussian disc}
\label{sec:model5}
In model 4, we noted that the three point sources can fit the excess \gray\ emission well. Considering that the two point sources, p2 and p3, are both more than $0.5\deg$ away from the best-fit position, thus we added these two point sources into model 5 based on the background model. Next, we added the centre of a Gaussian disc at the peak position of the residual map shown in Fig.\ref{fig:residual}. We assumed the Gaussian disc has a power-law spectral shape. The radius of the disc varies from $0.1\deg$ to $0.5\deg$ in steps of $0.01\deg$. The Gaussian disc template with a radius of $0.23\deg \pm\ 0.02\deg$ can best fit the \gray\ excess for the central part of RCW 38 with a $\rm TS_{ext}$ value of 130 listed in Table \ref{table:1}. The derived -$\log({\cal L})$ and AIC for this model are -3498124 and -6996074, respectively.

To compare the goodness of fit for the different models, we also calculated the $\rm \Delta AIC$, the AIC difference of model 1 and models 2-5. Table \ref{table:1} shows that the model 5 gives the highest $\rm TS_{ext}$ value and the minimum $\rm \Delta AIC$, thus the 0.23$\deg$ Gaussian disc and two point sources is the best-fit spatial template for the extended \gray\ emission.

\subsection{Spectral analyses}
\label{sec:spectral_analy}
\begin{figure}
\includegraphics[scale=0.41]{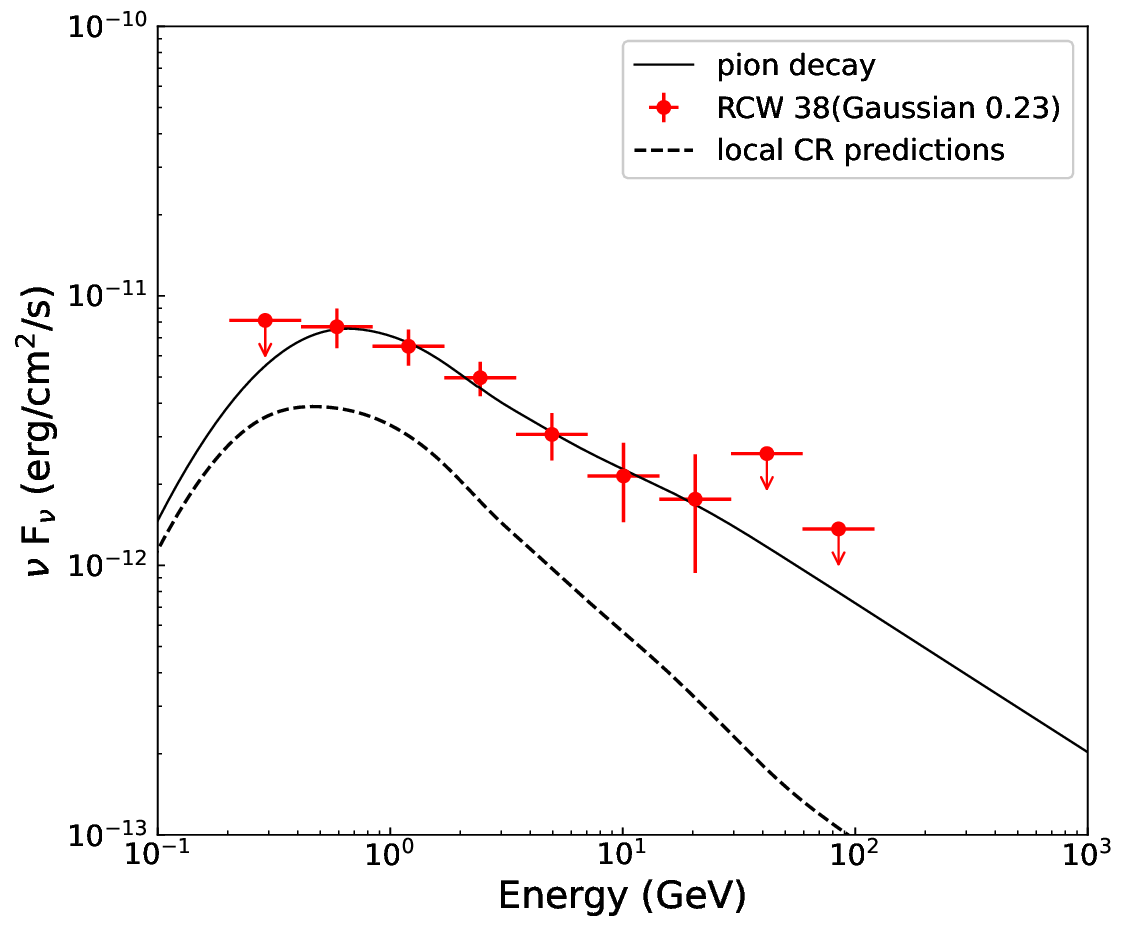}
\caption {SED of \gray\ emission toward RCW 38 for a spatially Gaussian disc model with a radius of $0.23\deg$. Both the statistical and systematic errors are considered. The dashed curve represents the predicted fluxes of \gray\ emission derived from the \ion{H}{II} and $\rm H_{2}$ column density map, the CRs are assumed to have the same spectra as measured locally (in the solar neighborhood) by AMS-02 \citep{2015Aguilar}. 
The solid curve represents the spectrum of \grays\ from interactions of relativistic protons with the ambient gas, assuming a power-law distribution of protons (See Sect.\ref{sec:origin}).}
\label{fig:sed}
\end{figure}

To find out the spectral shape of $0.23\deg$ Gaussian disc, we performed likelihood-ratio test for spectral models including LogParabola and BrokenPowerLaw, in which the PowerLaw (PL) model is the null hypothesis. The significance of the test model $\delta_{\rm model}$ is defined as $\sqrt{-2\log(\cal L_{\rm PL})/\log(\cal L_{\rm model})}$. We found that these spectral models do not improve the overall fitting results ($\delta_{\rm model}$ < 3), and the simple PL model is able to represent their spectral shape.
In this section we derived photon index for the $0.23\deg$ Gaussian disc with a single power-law spectrum  above 1 GeV is $2.44 \pm 0.03$ and the total \gray\ flux can be estimated as $(2.90 \pm 0.09) \times 10^{-9}\ \rm ph\ cm^{-2}\ s^{-1}$. 
Considering the distance of about 1.7 kpc and assuming the distance error is $10\%$ \citep{Peron:2023nu}, the total \gray\ luminosity is estimated to be $(1.61 \pm 0.05) \times 10^{33}\ \rm erg\ s^{-1}$. 
Then, We used the best-fit spatial template as the spatial model of the extended \gray\ emission and assumed a power-law spectral shape to extract the SED of the 0.23$\deg$ Gaussian disc. The derived SED is shown in Fig.\ref{fig:sed}.
We divided the energy range 200 MeV - 110 GeV into nine logarithmically spaced energy bins, and the SED flux in each bin is derived via the maximum-likelihood method. We calculated 95\% statistical errors for the energy flux densities.
We calculated the upper limits within 3$\sigma$ for the energy bins with a significance less than 2$\sigma$. In the analysis, we estimated the system uncertainties of the SEDs due to the Galactic diffuse emission model and the LAT effective area by changing the normalization by $\pm 6\%$ from the best-fit value for each energy bin, and considered the maximum flux deviations of the source as the systematic error \citep{Abdo09a}. The black dashed line in Fig.\ref{fig:sed} represents the predicted fluxes of \gray\ emissions based on the $\rm H_2$ + \ion{H}{II} column density map in this region, assuming that the CRs are the same as the local measurement by AMS-02 \citep{2015Aguilar}. Details for the gas content see Sect.\ref{sec:Gas}. The harder \gray\ spectrum is about 1-10 times than that of the local CRs. 

\section{Gas content around RCW 38}
\label{sec:Gas}

\begin{figure*}
\includegraphics[scale=0.24]{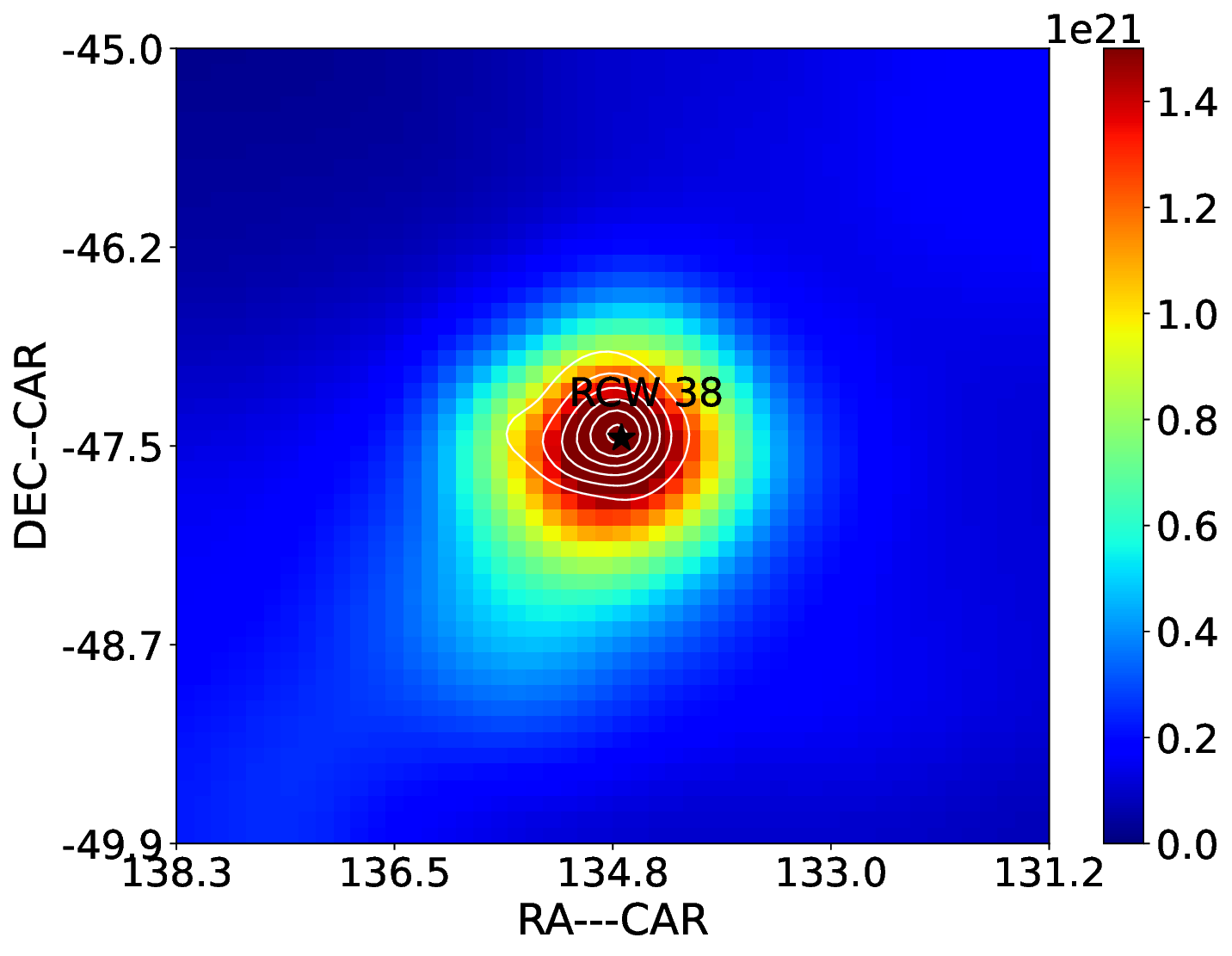}
\includegraphics[scale=0.24]{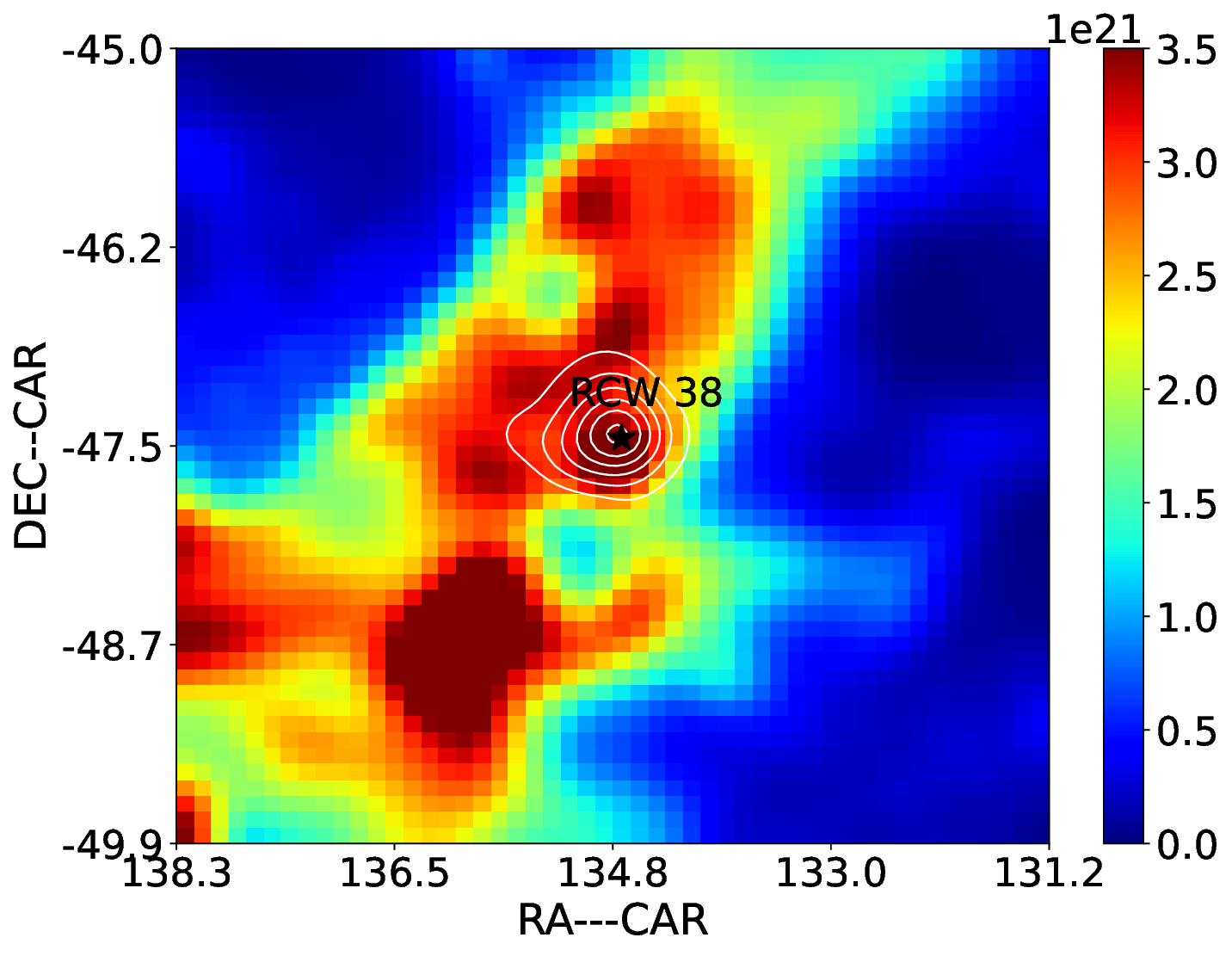}
\includegraphics[scale=0.24]{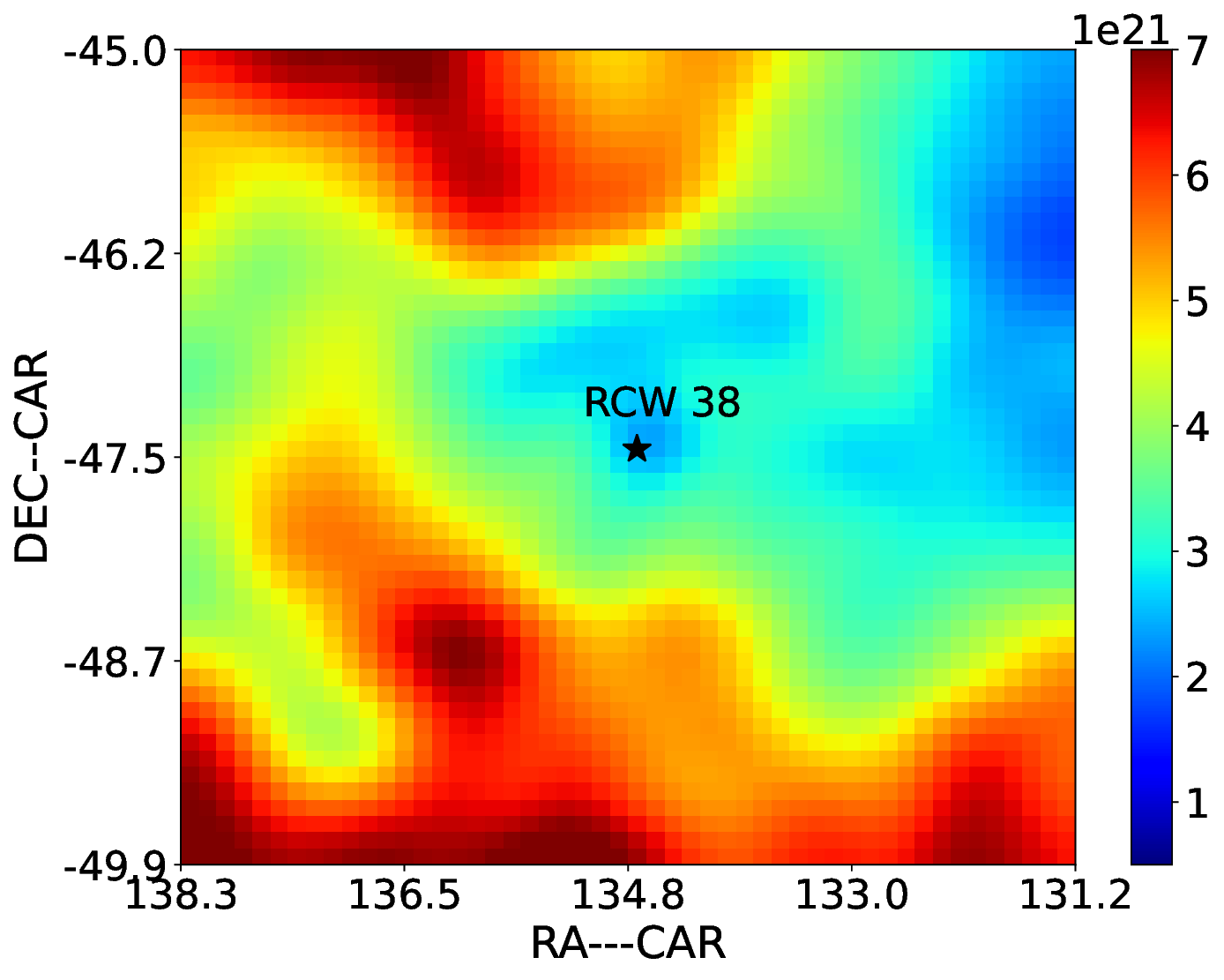}
\caption {Gas column densities (in units of $\rm cm^{-2}$) in three gas phases. The left panel shows the \ion{H}{ii} column density derived from the \planck\ free-free map assuming an effective density of electrons $n_{\rm e}=10~\rm cm^{-3}$. The middle panel shows the H$_{2}$ column density derived from CO data. The right panel shows the map of \ion{H}{i} column density derived from 21-cm all-sky survey. The black asterisk indicates the infrared position of RCW 38 \citep{2011Winston} and the white contours of the observed \gray\ emission derived in Sect.\ref{sec:model1} are overlaid in left and middle panels.
For details, see the context in Sect.\ref{sec:Gas}.}
\label{fig:Gas}
\end{figure*}

We investigated three different gas phases, i.e., the H$_{2}$, the neutral atomic hydrogen (\ion{H}{i}), and the \ion{H}{ii}, in the vicinity of the RCW 38 region.  
Radio surveys in the 1960s showed that RCW 38 to be one of the brightest \ion{H}{II} regions \citep{1970Wilson}.
To obtain the \ion{H}{ii} column density we used the \planck\ free-free map \citep{2016Planck}. 
First, we transformed the emission measure (EM) into the free-free intensity by using the conversion factor given in Table 1 of \cite{2003Finkbeiner}. 
Then, we calculate the \ion{H}{ii} column density from the intensity ($I_{\nu}$) of the free-free emission by using Eq.(5) of \cite{1997Sodroski}, 
\begin{equation}
\begin{aligned}
  N_{\ion{H}{ii}} = &1.2 \times 10^{15}\ {\rm cm^{-2}} \left(\frac{T_{\rm e}}{1\ \rm K}\right)^{0.35} \left(\frac{\nu}{1\ \rm GHz}\right)^{0.1}\left(\frac{n_{\rm e}}{1\ \rm cm^{-3}}\right)^{-1} \\
&\times \frac{I_{\nu}}{1\ \rm Jy\ sr^{-1}},
\end{aligned}
\end{equation}
where $\nu = \rm 353\ GHz$ is the frequency, and an electron temperature of $T_{e} =\rm 8000\ K$. 
The \ion{H}{ii} column density is inversely proportional to the effective electron density $n_{\rm e}$. 
Thus, we adopted an effective density of $10\ \rm cm^{-3}$, which is the value suggested in \cite{1997Sodroski} for the region inside the solar circle. 
The derived \ion{H}{ii} column density is shown in the left panel of Fig.\ref{fig:Gas}. 
We note that the \ion{H}{ii} gas distribution is consistent with the RCW 38 spatially, similar to that of the YMCs NGC 3603 \citep{2017Yang}, Westerlund 1 \citep{2012Abramowski}, Westerlund 2 \citep{2018Yang}, W40 \citep{W40Sun}, and Carina Nebula \citep{2022Ge}. The extended GeV \gray\ emissions with hard spectra have been detected from these clusters.
In addition, most of the observed \gray\ emission around these clusters shows good spatial consistency with the high \ion{H}{ii} column density region formed by the photoionization of these massive stars. 
Moreover, RCW 38 is the third YMC alongside Westerlund 2 and NGC 3603, where a cloud–cloud collision has triggered cluster formation, which is a possible sign of on-going O star formation \citep{2016Fukui}.

To track the distribution H$_{2}$ around the RCW 38 region, we use the CO composite survey \citep{2001Dame}. 
The standard assumption of a linear relationship between the velocity-integrated brightness temperature of the CO 2.6-mm line, $W_{\rm CO}$, and the column density of molecular hydrogen, $N(\rm H_{2})$, i.e., $N({\rm H_{2}}) = X_{\rm CO} \times W_{\rm CO}$ \citep{1983Lebrun}. 
$X_{\rm CO}$ is the $\rm H_{2} / CO$ conversion factor that chosen to be $\rm 2.0 \times 10^{20}\ cm^{-2}\ K^{-1}\ km^{-1}\ s$ as suggested by \cite{2001Dame} and \cite{2013Bolatto}. 
The derived molecular gas column density integrated in the velocity range $v_{\rm LSR}=[-8, 9]$ $\rm km\ s^{-1}$ \citep{2016Fukui} is shown in the middle panel of Fig.\ref{fig:Gas}. We also use this range to integrate the line emission of the \ion{H}{I} in this velocity range. As shown in Fig.\ref{fig:Gas}, the white contours represent the residual \gray\ emission towards RCW 38, which overlap well with the H$_{2}$ distribution from the CO measurements.

The \ion{H}{i} data are from the data-cube of the \ion{H}{i} $\rm{4\pi}$ survey (HI4PI), which is a 21-cm all-sky database of Galactic \ion{H}{i} \citep{HI4PI16}. 
We estimated the \ion{H}{i} column density using the equation,
\begin{equation}
N_{\ion{H}{i}} = -1.83 \times 10^{18}T_{\rm s}\int \mathrm{d}v\ {\rm ln} \left(1-\frac{T_{\rm B}}{T_{\rm s}-T_{\rm bg}}\right),
\end{equation}
where $T_{\rm bg} \approx 2.66\ \rm K$ is the brightness temperature of the cosmic microwave background radiation at 21 cm, and $T_{\rm B}$ is the brightness temperature of the \ion{H}{i} emission. 
In the case when $T_{\rm B} > T_{\rm s} - 5\ \rm K$, we truncate $T_{\rm B}$ to $T_{\rm s} - 5\ \rm K$; $T_{s}$ is chosen to be 150 K. 
The derived \ion{H}{I} column map is shown in the right panel of Fig.\ref{fig:Gas}. The \ion{H}{I} column density shows spatial inconsistency with the observed \gray\ emissions. Therefore, the \ion{H}{I} gas is not considered in the following calculations.

\begin{table}
        \caption{Different gas masses and number densities within the Gaussian disc with a radius of $0.23\deg$. See Sect.\ref{sec:Gas} for details.}
\begin{tabular}{lcccc}
\hline
        Tracer & Mass ($\rm{10^{2}\msun}$) & Number density ($\rm {cm^{-3}}$) \\
\hline
        \ion{H}{II} & 89.89 & 67 \\ 
        H$_{2}$ & 349.65 & 261 \\ 
        \ion{H}{II} + H$_{2}$ & 439.54 & 328 \\ 
\hline
\end{tabular}\\
\label{table:2}
\end{table}

The total mass within the cloud in each pixel can be calculated from the expression
\begin{equation}
M_{\rm H} = m_{\rm H} N_{\rm H} A_{\rm angular} d^{2}
\end{equation}
where $M_{\rm H}$ is the mass of the hydrogen atom, $N_{\rm H} = N_{\ion{H}{ii}} + 2N_{\rm H_{2}}$ is the total column density of the hydrogen atom in each pixel. $A_{\rm angular}$ is the angular area, and $d$ is the distance of RCW 38. 
We calculated the total mass and number of hydrogen atoms in each pixel. The total mass in the GeV \gray\ emission region is estimated to be $\sim 4.4 \times 10^{4}~\msun$ as listed in the Table \ref{table:2}. 
If we assume that the GeV \gray\ emission within the region is spherical in geometry, with the corresponding size of $0.23\deg$. For a symmetric 2D Gaussian distribution $\theta_{95}$ is $1.62 \times \theta_{68}$, here $\theta_{95}$ and $\theta_{68}$ are the radian within 95\% and 68\%, respectively.
The radius can be estimated as $r = d \times \theta (\rm{rad}) \sim 1700\ \rm pc \times \ (0.23\deg \times 1.62 \times \pi / 180\deg) \sim 11\ \rm pc $, and $d$ is the distance to the objective region. 
The total gas number density averaged over the volume of the \gray\ emission region is $\rm n_{gas}=328\ cm^{-3}$.
Table \ref{table:2} shows the different gas masses and number densities within the \gray\ emission region of RCW 38.

\section{The origin of gamma-ray emission}
\label{sec:origin}
To investigate the possible radiation mechanisms of the \grays\ in the RCW 38 region, we fit the SED of the $0.23\deg$ Gaussian disc with both leptonic and hadronic scenarios. We used Naima \footnote{\url{https://naima.readthedocs.io/en/latest/index.html}} \citep{2015Zabalza} to fit the SEDs. Naima is a numerical package that allows us to implement different functions and includes tools to perform Markov Chain Monte Carlo (MCMC) fitting of non-thermal radiative processes to the data.

\subsection{Hadronic Scenario}
%
The GeV \gray\ emission around RCW 38 is spatially consistent with the high density molecular and ionized hydrogen gas distribution. The consistency supports the hadronic origin, that is, the emission comes from the decay of neutral pions produced by the interactions between the accelerated hadrons and the surrounding gas. 
Thus, we assume that the \grays\ emission are produced in the pion-decay process from the interaction of the CRs with the ambient clouds. The average number density of the target protons for this region is $328\  \rm cm^{-3}$ derived from the gas distributions in Sect.\ref{sec:Gas} considering $\rm H_2$ + \ion{H}{II} gas. We used a single power-law spectrum for the parent proton distribution,
\begin{equation}
        N(E) = A~E^{-\alpha},
\label{equ:pl}
\end{equation} 
treating $A$, $\alpha$ as free parameters for the fitting. 
As shown in Fig.\ref{fig:sed}, we present the best-fit results for the SED of Gaussian disc with a radius of $0.23\deg$. The maximum log-likelihood value is -0.21.
The derived index of this region, $\alpha = 2.62 \pm 0.06$, with the total energy $W_{\rm p} = (2.1 \pm 0.06) \times 10^{47}\ \rm erg$ for the protons above 2 GeV. 
Here, we consider whether the massive OB stars in RCW 38 cluster could be the origins as an example. The kinematic luminosity $L_{w}$ that can be supplied by the wind from a single massive star is obtained from the following
\begin{equation}
L_{w} = \frac{1}{2} \dot{M} v_{w}^{2}= 1\times10^{35} {\rm erg\ s^{-1}} (\frac{\dot{M}}{10^{-7}\ {\rm M_{\odot}}\ {\rm yr^{-1}}})(\frac{v_{w}}{2000\ {\rm km\ s^{-1}}})^2,  
\label{equ:Lw}
\end{equation}
where $\dot{M} = 10^{-7} M_{\odot} \rm yr^{-1}$ is the standard mass loss rate \citep{2001Vink} and $v_{w} = 2000\ {\rm km\ s^{-1}}$ is the typical wind velocity \citep{1990Prinja}. Since more than 30 OB stars have been reported to be associated with RCW 38 \citep{2006Wolk}, there should be an energy supply of more than $3 \times 10^{36}\rm erg\ s^{-1}$. Taking into account the age of 1 Myr, the total energy injected by the RCW 38 cluster is about $9.5 \times 10^{49} \rm erg$. Thus, this system is powerful enough to accelerate enough CRs to account for the detected \gray\ emissions. 


\begin{figure}
\includegraphics[scale=0.41]{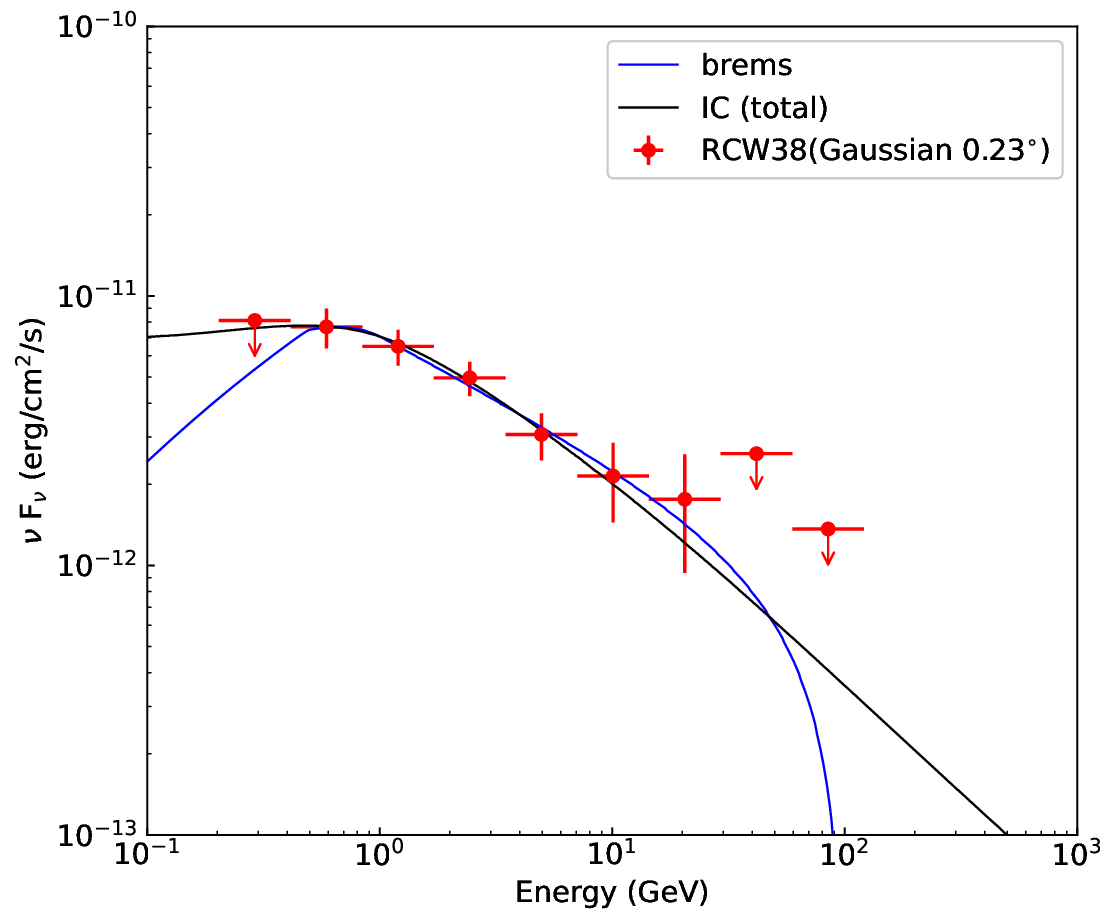}
\caption {Same as Fig.\ref{fig:sed} but for leptonic modeling. The black solid curve represents the spectrum of \grays\ from the IC process. The blue solid curve shows the corresponding bremsstrahlung spectrum for an ambient matter density of $328\ \rm cm^{-3}$. For details, see the context in Sect.\ref{sec:origin}.}
\label{fig:sed_IC}
\end{figure}

\subsection{Leptonic Scenario}
We considered the leptonic scenario, where the GeV \gray\ emission are generated via the inverse Compton (IC) scattering of relativistic electrons from the low-energy seed photons, or through nonthermal bremsstrahlung from the relativistic electrons or matter around the RCW 38 region. 
For the photon field of the IC calculations, we considered the cosmic microwave background (CMB) radiation field, optical-UV radiation field from the star light, and the dust infrared radiation field based on the model by \cite{2017Popescu}. 
In this case, RCW 38 is located in the \ion{H}{II} region, the ionizing massive stars will increase the optical and UV fields significantly and thus produce additional IC emissions \citep{2022Liu&Yang}. 
The target particle density is assumed to be $328\ \rm cm^{-3}$ for the relativistic bremsstrahlung, noting that larger or smaller values of the ambient matter density simply scale the contribution of the bremsstrahlung spectrum. 
We calculated the IC spectrum using the formalism described in \cite{2014Khangulyan}, and for the relativistic bremsstrahlung spectrum we used the parameterization in \cite{1999Baring}. To fit the lower energy break in the \gray\ spectrum, we should require a relevant break in the spectrum of parent electrons. Thus, we assumed a broken power-law distribution of the relativistic electrons,
\begin{equation}
N(E) =\begin{cases} A(E/E_0)^{-\alpha_1} & \mbox{: }E<E_\mathrm{b} \\ A(E_\mathrm{b}/E_0)^{(\alpha_2-\alpha_1)}(E/E_0)^{-\alpha_2} & \mbox{: }E>E_\mathrm{b} \end{cases},
\label{equ:bpl}
\end{equation}
treating A, $\alpha_1$, $\alpha_2$, $E_\mathrm{b}$ as free parameters for the fitting. 
As is shown in Fig.\ref{fig:sed_IC}, both the IC and bremsstrahlung model can fit the observable data. The corresponding maximum-likelihood values is -2.51 and -0.24, respectively. For IC model, the derived parameters for the electrons are $\alpha_1 = 0.6 \pm 0.2$, $\alpha_2 = 4.1^{+0.6}_{-0.3}$, $E_\mathrm{b} = 10.7 \pm 1.2\ \rm GeV$, and the total energy of the electrons above 2 GeV is $W_{\rm e} = (2.2 \pm 0.3) \times 10^{49}\ \rm erg$. 
For bremsstrahlung, $\alpha_1 = 0.76 \pm 0.11$, $\alpha_2 = 2.62^{+0.13}_{-0.09}$, $E_\mathrm{b} = 1.01 \pm 0.15\ \rm GeV$, and $W_{\rm e} = (9.7 \pm 1.1) \times 10^{45}\ \rm erg$. 
We can not formally rule out a leptonic origin for this source. In such a dense region, bremsstrahlung dominates the radiation mechanism. The good spatial coincidence of the \gray\ emission with the \ion{H}{II} gas region favours a hadronic or bremsstrahlung origin. However, in the bremsstrahlung scenario, the derived parent proton index changes from 0.76 to 2.62 below and above about 1 GeV. Such a sharp break is quite unusual and incompatible with any known mechanism.

\section{Discussion}
\label{sec:Disc}
We find two pulsars PSR J0855-4644 and PSR J0855-4658 within 1$\deg$ of the peak of the \gray\ emission from the Australia Telescope National Facility Pulsar Catalogue \citep{2005Manchester}.
Their spin-down luminosities are $1.1\times 10^{36}\ {\rm erg\ s^{-1}}$ and $2.8\times 10^{33}\ {\rm erg\ s^{-1}}$, while their distances are 9.9 kpc and 28.3 kpc, respectively \citep{2003Kramer}. PSR J0855-4658 has a relatively low spin-down power, so we consider it's an unlikely source of the \grays\ emission. \xray\ observations have identified a pulsar wind nebula (PWN) in the region around PSR J0855-4644, which coincides with the shell of RX J0852.0-4622 \citep{2013Acero}. This pulsar is energetic enough to power a very high-energy PWN that is detectable by current generation Cherenkov telescopes \cite{2018HESSpulsar}. We cannot confirm or rule out this source as the origin of the GeV \gray\ emission, but they are both more than $\sim 0.8\deg$ away from the peak of the \gray\ emission, making this scenario quite unlikely.
In the northwestern part of the RCW 38 cluster, there is a known SNRs Vela Jr. (RX J0852.0-4622, see Fig.\ref{fig:cmap}), which is more than $\sim$ 1\deg away from RCW 38. Vela Jr. is a young shell-type SNR detected at very high energy \citep{2018HESS}. This SNR is located at a distance of $\sim$700 pc ($\pm$ 200 pc) \citep{2015Allen}, which is an incompatible distance with RCW 38. 

Another scenario is to associate the extended GeV \gray\ emission with the YMC RCW 38. Indeed, the spectral shape and spatial extension are similar to those measured in other YMCs. The total CR energy in RCW 38 is only to the order of $10^{47}$ erg, which is much lower than that derived from similar systems such as NGC 3603 \citep{2017Yang}, Westerlund 2 \citep{2018Yang}, and the Cygnus cocoon \citep{2011Ackermann, 2019Aharonian}. Also, RCW 38 is much less powerful than the other detected systems. Comparing with other systems, there are about 30 identified OB stars and the cluster mass is less than $10^{4} \msun$. Therefore, it is reasonable to assume that the wind power in the RCW 38 cluster is orders of magnitude lower. This may also be due to the younger age of this star cluster. Thus, if \grays\ are illuminated by CRs accelerated in RCW 38, the natural acceleration sites of the CRs are the stellar winds of young massive stars. The derived $W_{\rm p}$ is $2.1 \times 10^{47}\ {\rm erg}$ for the protons above 2 GeV, considering the wind power of the YMC RCW 38 of $3\times 10^{36}\ {\rm erg\ s^{-1}}$ and an acceleration efficiency of 10 per cent. The CR injection power can be $P_{\rm CR} \sim 3\times 10^{35} ~\rm erg/s$. Taking into account the size of this region of about $l \sim 11 \rm pc$ ($0.23\deg$ Gaussian disc in $1.7~\rm kpc$, see Sect.\ref{sec:Gas}), the required diffusion coefficient in this region can be estimated as $D \sim \frac{l^2}{4T} $, where the confinement time $T$ can be estimated as $W_{\rm p}/P_{\rm CR} \sim 7\times 10^{11} \rm s$. The derived $D$ is $4 \times 10^{26} ~\rm cm/s$, two orders of magnitude smaller than the average value in the  galactic plane. CRs produced in RCW 38 may be confined inside the source due to slower diffusion, which forms the \gray\ emission region. 

\section{Conclusion}
\label{sec:Conc}
In this paper, we report the detection of GeV \gray\ emission toward the RCW 38 region, which is a YMC in our Galaxy. We found that the extended \gray\ emission can be modelled by a Gaussian disc with a radius of $0.23\deg$. Like the other YMCs, the \gray\ emission reveals a hard spectrum which can be described by a power-law function with a photon index of about $2.44 \pm 0.03$. Compared to the result obtained by \cite{Peron:2023nu}, the result of GeV \gray\ emission around the RCW 38 cluster is consistent with their result. The harder \gray\ spectrum is about 1-10 times higher than that of the local CRs. Due to the high gas density in this region, we argue that the most likely origin of the detected GeV \gray\ emission is the interaction of CRs accelerated in the young star cluster with the ambient gas. Compared with other systems, RCW 38 is extremely young at less than 1 Myr. Within such a young star forming system, practically no star has ever evolved to explode as a supernova. Additionally, the centre of the detected \gray\ emission is 0.015\deg away from the stellar cluster itself. Thus, it is likely that the observed \gray\ emission is directly from the stellar cluster, similar to that of the YMC W40 \citep{W40Sun}. Since YMCs are likely to be another CR source, the propagation of CRs in the vicinity of these sources will be the key to understand the injection of CRs, which can be verified by better spatial and energy spectral information in the future.

\section{Acknowledgements}
We thanks Yun-Feng Liang for useful discussion. This work is supported by the National Natural Science Foundation of China (NSFC, Grant No.12133003, 12103011, and U1731239), Science and Technology Program of Guangxi (Grant No.AD21220075), Innovation Project of Guangxi Graduate Education (Grant No.). Rui-zhi Yang is supported by NSFC under grant 12041305 and the national youth thousand talents program in China. P.H.T.T. is supported by NSFC grant 12273122 and a science research grant from the China Manned Space Project (No. CMS-CSST-2021-B11).

\section{Data availability}
The \fermi\ data used in this work are publicly available, and are provided online at the NASA-GSFC Fermi Science Support Center\footnote{\url{ https://fermi.gsfc.nasa.gov/ssc/data/access/lat/}}.
We made use of the CO data\footnote{\url{ https://lambda.gsfc.nasa.gov/product/}} to derive the H$_{2}$ column density. The data from \planck\ legacy archive\footnote{\url{ http://pla.esac.esa.int/pla/\#home}} were used to derive the \ion{H}{ii} column density.
The \ion{H}{i} data were taken from the HI4PI\footnote{\url{http://cdsarc.u-strasbg.fr/viz-bin/qcat?J/A+A/594/A116}}.

\bibliographystyle{mnras}
\bibliography{ms}

\bsp	
\label{lastpage}
\end{document}